\documentclass[pra,twocolumn,showpacs]{revtex4}

\usepackage{epsfig,amsmath}

\def\sref#1{Sec.~(\ref{#1})}
\def\fig#1{Fig.~\ref{#1}}

\begin{document}

\title{Disorder Correlation Frequency Controlled Diffusion in the Jaynes-Cummings-Hubbard Model}

\author{James Q. Quach} 
\email{quach.james@gmail.com}
\affiliation{School of Physics, The University of Melbourne, Victoria 3010, Australia}

\begin{abstract}
We investigate time-dependent stochastic disorder in the one-dimensional Jaynes-Cummings-Hubbard model and show that it gives rise to diffusive behaviour.  We find that disorder correlation frequency is effective in controlling the level of diffusivity. In the defectless system the mean squared displacement (MSD), which is a measure of the diffusivity, increases with increasing disorder frequency.  Contrastingly, when static defects are present the MSD increases with disorder frequency only at lower frequencies; at higher frequencies, increasing disorder frequency actually reduces the MSD.
\end{abstract}


\maketitle

\section{Introduction}

\begin{figure}[tb]%
	\centering
	\includegraphics[width=\columnwidth]{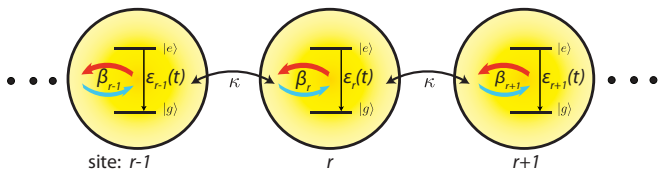}%
\caption{Schematic of the 1D JCH model. At each site, a two-level atom interacts with the quantized cavity field with coupling strength $\beta$. The ground $|g\rangle$ and excited $|e\rangle$ states of the atom are separated by atomic transition energy $\epsilon$. Post fabrication control of $\epsilon(t)$ provides the dynamic disorder. These individual JC cavities are coupled together into an array with intercavity coupling given by the hopping frequency, $\kappa$.}%
\label{fig:jch_1D}%
\end{figure}

The interaction of an optical cavity with a two-level atom is described by the well known Jaynes-Cummings (JC) model~\cite{jaynes63}. The coupling of these JC systems in a tight-binding model is known as the Jaynes-Cummings-Hubbard (JCH) (\fig{fig:jch_1D}) model~\cite{hartmann06,greentree06,angelakis07}. The JCH model has been shown to be a rich dynamical platform, giving rise to versatile properties. It has been proposed as a good candidate structure on which to build quantum emulators~\cite{greentree06,quach09,hayward12} and quantum metamaterials~\cite{quach11}. Here we will show for the first time, controlled diffusion in the JCH model. Diffusion as a tunable parameter in JCH systems will significantly enhance its capability as a quantum metamaterial and also as a quantum emulator of, for example, quantum Brownian motion.

Thermal agitation has been shown to give rise to diffusion in a non-interacting tight-binding lattice~\cite{madhukar77}. The thermal agitation was modelled as a random disorder, uncorrelated in space and time, where the mean of the disorder vanishes.  The level of diffusion was determined by the strength of the spatial disorder. An alternative controlling parameter, which has not been investigated in lattice models, is the frequency of the disorder. Here we will show how disorder frequency controls diffusion in the single-excitation subspace of the JCH model. 

The strong coupling of atom or atom-like systems to microcavities, which form the unit elements of JCH systems, has been achieved in a variety of designs including: cesium beams intersecting the cavity axis of Fabry-Perot (FP) resonators~\cite{rempe91,thompson92}; cesium atoms in microtoroids~\cite{aoki06}; single rubidium atoms in FP resonators~\cite{nussmann05,hijlkema2007single}; quantum dots in FP-based micropillars~\cite{reithmaier04,press07}, photonic crystals (PhCs)~\cite{yoshie04,englund05,hennessy07}, and microdisks~\cite{peter05,srinivasan07}; and diamond defects in PhCs~\cite{barth09}, microdisks~\cite{barclay09a,barclay2009coherent}, and microspheres~\cite{park06}. Circuit QED provides another viable alternative. In this design on-chip superconducting coplanar waveguide microwave resonators serves as the effective cavity~\cite{goppl08} with Cooper-pair boxes~\cite{wallraff04}, transmon\cite{koch2007charge} or flux qubits~\cite{lindström2007circuit} acting as the artificial atom. 

A number of small scale (12 to 30) coupled microcavities have been constructed with microrings~\cite{poon2006transmission}, microspheres~\cite{hara2005heavy}, and PhCs~\cite{o2007coupled}. Large scale arrays of over 100 cavities have been demonstrated with silicon ring microcavity based coupled resonator optical waveguides~\cite{xia2006ultracompact} and PhC designs~\cite{notomi08}. The coupling of microcavities with atomic systems into large arrays, which is what is needed to realise a useful JCH system, is yet to be experimentally demonstrated. However designs of one-dimensional (1D) arrays of waveguide-coupled atom-optical cavities~\cite{lepert2011arrays} have been proposed. Diamond photonic bandgap structures with nitrogen-vacancy (NV) centres in one~\cite{makin09, quach09} and two~\cite{greentree06} dimensions have also been put forth as possible designs. With the rate of recent advances in microcavity fabrication technology, it is foreseeable that large scale coupled-cavities with atomic systems will be the next advancement in experimental development.

Currently and in the foreseeable future, one of the main technical challenges in the physical realisation of the JCH model is the efficient fabrication of precise unit elements. Some technical challenges exist in the fabrication of uniform sets of microcavities and constructions of uniform microcavity arrays. Even more challenging is fabricating consistent atom-cavity couplings. As defects are likely to be common, it is important to investigate their influence on the dynamics of JCH systems. In this work we will investigate diffusion in the ideal defectless system  as well as in the presence of defects.

This paper is organised as follows. \sref{sec:Model} introduces the JCH Hamiltonian with dynamic stochastic disorder. In \sref{sec:Frequency Controlled Diffusion} we investigate dynamic disorder in the ideal defectless system, i.e. when there is no static disorder. We show how disorder frequency can control the rate of diffusion. In \sref{sec:Diffusion with Static Disorder} we investigate dynamic disorder in systems with defects i.e. static disorder. We show the influence the presence of defects have on diffusive behaviour.

\section{Model}
\label{sec:Model}

The JC Hamiltonian is given by ($\hbar=1$),
\begin{equation}
	H^{\rm{JC}} = \epsilon \sigma^+ \sigma^- + \omega a^\dagger a + \beta(\sigma^+ a + \sigma^- a^\dagger),
\label{eq:jaynes_cummings_Hamiltonian_1}
\end{equation}
where $\omega$ is the resonant cavity frequency and $a(a^\dagger)$ is the photonic lowering (raising) operator. $\epsilon$ is the atomic transition energy and $\sigma (\sigma^\dagger)$ is the atomic lowering (raising) operator. $\beta$ is the atom-cavity coupling strength. The JCH Hamiltonian describes the interaction of coupled JC systems,
\begin{equation}
	\mathcal{H} = \sum_r {H_r^{\rm{JC}}} - \sum_{\langle r, s \rangle}{\kappa_{rs} a_r^\dagger a_s},
\label{eq:jch}
\end{equation}
where $r$ and $s$ are cavity site indices and $\kappa_{rs}$ is the hopping frequency between cavity $r$ and $s$. 

A viable means to dynamically vary the properties of the JCH systems post fabrication, is to tune the atomic transition energy.  This can be achieved by applying a controlled external electric field to induce the Stark effect. In particular, Stark shift control of NV centers have been  experimentally demonstrated with step-wise application of electric fields~\cite{tamarat06}. Here we consider the case where the atomic transition energy is a controlled stochastic variable which is uncorrelated in space but correlated in time,
\begin{equation}
	\epsilon_r(t)=\alpha + \gamma_r(t)~,
\label{eq:epsilon}
\end{equation}
where $\alpha$ is constant and $\gamma_r(t)$ is a Gaussian random variable with a spectrum specified by,
\begin{equation}
\begin{split}
	\langle\gamma_r(t)\rangle&=0~,\\
	\langle\gamma_r(t)\gamma_{r'}(t')\rangle&=p\delta_{rr'}[h(t-t')-h(t-t'-\tau)]~.
\end{split}
\label{eq:alpha}
\end{equation}
$p$ is a measure of the disorder strength, $\delta_{rr'}$ is the kronecker delta, $h$ is the Heaviside step function, and $\tau$ is a finite correlation time, which is the inverse of the disorder correlation frequency (DCF), $f_\text{D}=1/\tau$. 

\section{Frequency Controlled Dispersion-Diffusion Transition}
\label{sec:Frequency Controlled Diffusion}

We consider a 1D JCH chain in the single-excitation subspace with no defects, i.e.  $\omega$, $\beta$, $\kappa$, and $\alpha$ are uniform. We operate in the strong atom-cavity coupling regime, not too far from resonance: $\beta=10\kappa$, $\alpha=\omega$, $|\gamma_r(t)| \le 0.2\beta$. Note that the absolute value of $\omega$ here is arbitrary, as the dynamics of the system is dependent on its relative value to $\epsilon$, i.e. $\epsilon - \omega = \gamma$.

\begin{figure*}[tb]%
  \center
	\includegraphics[width=2\columnwidth]{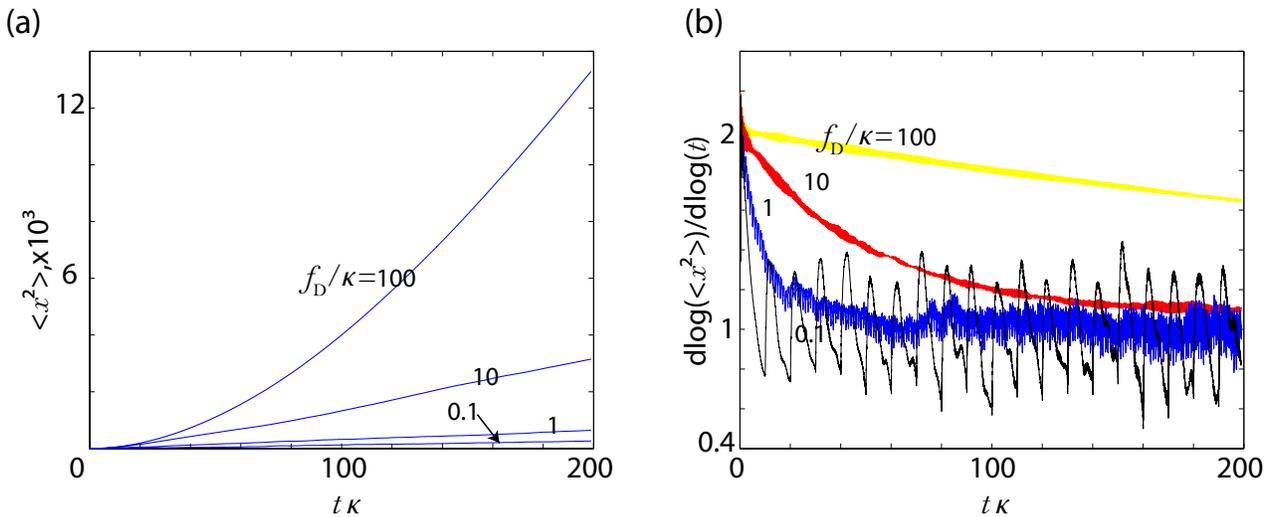}%
\caption{(color online) (a) MSD averaged over 500 samples in a defectless system with dynamic disorder for various DCFs, $f_\text{D}$. (b) $d\log(\langle x^2 (t)\rangle)/d\log(t)$ averaged over 500 samples. System is initially dispersive ($\langle x^2(t)\rangle \sim t^2$) before graduating to a more diffusive regime at later times ($\langle x^2 (t)\rangle \sim t$).}%
\label{fig:MSDvsT_defectless}%
\end{figure*}

Let $|g, 1\rangle$ and $|e, 0\rangle$ represent the photonic and atomic modes of the atom-cavity system in the single excitation subspace. We initialise the system in a state with excitation localised to a single site $r_0$, equally distributed between the atomic and photonic modes, $|\psi(r_0,t)\rangle = (|g,1\rangle + |e,0\rangle/\sqrt{2})$. The propagation of the field in the lattice is governed by the Schr\"{o}dinger equation $|\psi(x,t)\rangle = e^{i \mathcal{H}t} |\psi(x,0)\rangle$. As the lattice size used in each simulation is chosen to be much larger than the time evolved spatial distributions of the excitation, the influence of the lattice boundaries can be ignored. Under this condition varying the lattice size has negligible effect.

To see how the excitation spatially spreads over time, we calculate the mean square displacement (MSD), $\langle x^2(t) \rangle = \sum_r[|\psi(r-r_0,t)|(r-r_0)]^2$. \fig{fig:MSDvsT_defectless}(a) plots the MSD behaviour for various DCF, averaged over 500 samples. \fig{fig:MSDvsT_defectless}(b) plots $d\log(\langle x^2 (t)\rangle)/d\log(t)$ averaged over 500 samples, which gives the degree of a monomial MSD. Ref.~\cite{madhukar77} showed that for temporally uncorrelated disorder in a non-interacting tight-binding lattice, the MSD behaviour is initially dispersive ($\langle x^2(t)\rangle \sim t^2$) before graduating to a more diffusive regime at later times ($\langle x^2 (t)\rangle \sim t$). \fig{fig:MSDvsT_defectless}(a) and (b) indicate that we have similar behaviour here. 

The simulation results show that the system initially behaves dispersively, irrespective of the DCF. However, the rate at which the system transits from dispersive to diffusive behaviour is dependent on the DCF. \fig{fig:MSDvsT_defectless}(b) shows that the rate of transition from dispersion to diffusion decreases with DCF. On the timescale of the simulations, the system has transited to an approximately diffusive regime for $f_\text{D}/\kappa=0.1,1,10$, but the slower rate of transition for $f_\text{D}/\kappa=100$ means that this system is still relatively dispersive. The gradient of the plots in \fig{fig:MSDvsT_defectless}(a) also shows that in the diffusive regime, the level of diffusivity increases with DCF.

The transition from dispersive to diffusive behaviour is not a smooth one. During the autocorrelated time periods $[t_i,t_i+\tau]$, there is a decrease in the rate of change of the MSD due to Anderson localisation. However, the abrupt change in stochastic disorder at each uncorrelated time instant $t_i$ gives rise to temporary delocalisation, which produces a surge in the rate of change of the MSD, before localisation sets in again in the autocorrelated period. Although this high-frequency oscillatory behaviour with period $\tau$ is most apparent for $f_\text{D}/\kappa=0.1$ in  \fig{fig:MSDvsT_defectless}(b), the other simulations also undergo the same dynamics but with smaller amplitudes.

\begin{figure*}[tb]%
  \center
	\includegraphics[width=2\columnwidth]{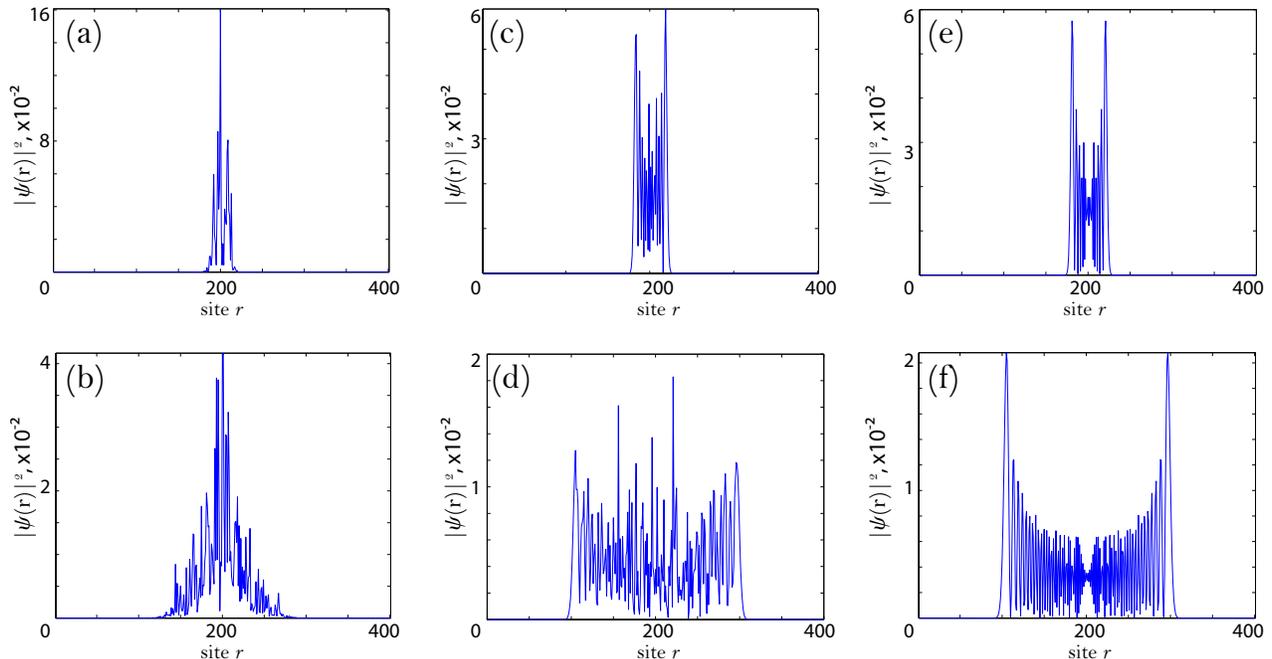}%
\caption{Time snapshots of a defectless system with dynamic disorder. At $t=0$, the initial state is localised to site $r=200$, and time snapshots of the population distribution is taken at $t\kappa=10$ [(a),(c),(e)] and $t\kappa=100$ [(b),(d),(f)] for various DCFs. (a)-(b)$f_\text{D}=\kappa$, the system distributes more diffusively. (c)-(d) $f_\text{D}=10\kappa$, the system distributes more dispersively. As a comparison, (e)-(f) shows time snapshots of the purely dispersive behaviour of a defectless system without any disorder ($\gamma_r = 0$).}
\label{fig:states}%
\end{figure*}

The difference in dispersive and diffusive behaviour is further illustrated in \fig{fig:states}, where the spatial distributions of the excitation for different DCF and time instances are shown. At t=0, the initial state is localised to site $r=200$.  \fig{fig:states}(a),(b) shows the time snapshots of the population distribution $|\psi(r)|^2$ at $t=10/\kappa$ and $t=100/\kappa$ for the case when $f_\text{D}=\kappa$. \fig{fig:states}(c),(d) shows the same time snapshots for $f_\text{D}=10\kappa$. As a comparison \fig{fig:states}(e),(f) contains the time snapshots of an ideal system without any dynamic disorder ($\gamma_r=0$). \fig{fig:states}(a)-(f) shows that for $f_\text{D}=\kappa$ the system distributes more diffusively, whereas for $f_\text{D}=10\kappa$ the system behaves more dispersively at this timescale, i.e. its behaviour is closer to the ideal dispersive system with no dynamic disorder. 

To understand why the MSD rate increases with DCF, first consider the case when $f_\text{D}=0$. The static stochastic disorder yields spatially non-overlapping sets of eigenstates, which leads to Anderson localisation. However when the DCF is finite, the sets of eigenstates which are localised in space can effectively overlap in time. Let us be specific about what we mean by this. Let $\Psi(t)$ be the localised sets of eigenstates, which changes at each uncorrelated time instant. Let $\Phi_1\in\Psi(t_1)$ be localised sets of eigenstates that are populated at time $t_1$. At the next uncorrelated time instant, $t_2 = t_1 + \tau$, there is a change in the eigenstates of the system, and the populated localised sets of eigenstates $\Phi_2$, are the eigenstates accessible from $\Phi_1$ which is in $\Psi(t_2)$. In general, $|\Phi_2|\ge|\Phi_1|$, where $|\Phi|$ means the cardinality of $\Phi$.  The rate at which the cardinality of the populated eigenstates grows, which is a measure of rate of increase of the MSD, increases with the DCF. This also explains why we see a surge in $d\log(\langle x^2 (t)\rangle)/d\log(t)$ at each sudden change in the stochastic disorder: at each uncorrelated time instant, there is an instantaneous increase in the cardinality of $|\Phi|$, allowing the population to quickly spread.

\section{MSD behaviour with Static Disorder}
\label{sec:Diffusion with Static Disorder}

Defects are likely to arise in the fabrication of the quantum components that comprise the JCH system. In particular, there are great technical challenges in producing a system with a uniform atom-coupling constant, as $\beta$ is highly dependent on the relative location of the atom in the cavity. In this section we look at the diffusion in systems with defects. Specifically we consider the uncorrelated disorder case, i.e. $\beta_r$ is a spatially uncorrelated time-independent stochastic variable.

\begin{figure*}[tb]%
  \center
	\includegraphics[width=2\columnwidth]{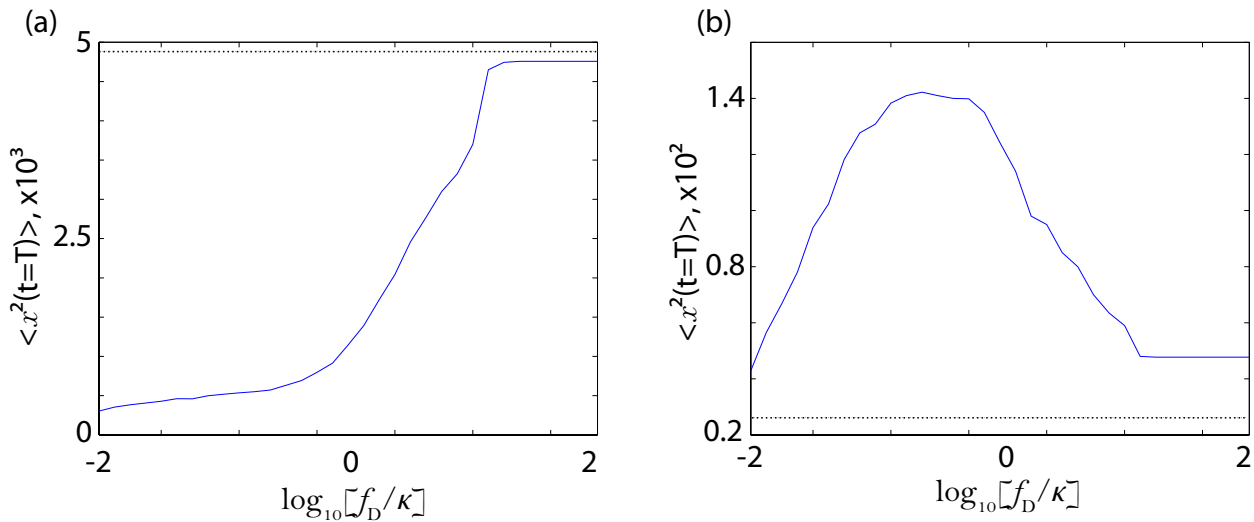}%
\caption{MSD at $T=100/\kappa$ averaged over 500 samples, as a function of DCF. $\beta_r$ gives the static disorder and $\epsilon_r(t)$ provides the dynamic disorder. (a)  When there are no defects ($\beta$ is uniform), MSD increases with DCF, as expected. (b) When there are static defects ($\beta_r$  is a spatially stochastic variable), MSD increases with DCF at lower frequencies, but as the frequencies get higher, MSD actually decreases as the system increasingly acts like its time averaged behaviour, and localisation occurs again. Dotted lines indicates the MSD of the respective systems with no dynamic disorder. }%
\label{fig:rmsVSfreq}%
\end{figure*}

We conduct the simulation of \sref{sec:Frequency Controlled Diffusion}, but now for both uniform $\beta$ [\fig{fig:rmsVSfreq}(a)] and when $\beta_r$ [\fig{fig:rmsVSfreq}(b)] is a time-invariant spatially uncorrelated stochastic parameter,

\begin{equation}
	\beta_r=\zeta + \eta_r~,
\label{eq:beta}
\end{equation}
where $\zeta$ is constant and $\eta_r$ is a Gaussian random variable with a spectrum specified by,
\begin{equation}
\begin{split}
	\langle\eta_r\rangle&=0~,\\
	\langle\eta_r\eta_{r'}\rangle&=p\delta_{rr'}~.
\end{split}
\label{eq:eta}
\end{equation}

As with \sref{sec:Frequency Controlled Diffusion} $\epsilon_r(t)$ provides the temporal stochastic disorder, and we work in the strong coupling regime and near resonance: $\zeta = 10\kappa$, $\alpha=\omega$ , $|\gamma_r(t)|\le 0.1\zeta$, $|\eta_r| \le 0.1\zeta$. We plot the MSD at time $T=100/\kappa$, averaged over 500 samples, as DCF is varied. The lattice size used is large relative to the spatial distribution of the excitation at $T=100/\kappa$ so that boundary effects are negligible.

\fig{fig:rmsVSfreq}(a) shows that for uniform $\beta=\zeta$, MSD increases with DCF, as expected. Contrastingly, when there are defects and $\beta_r$ is a spatially stochastic variable [\fig{fig:rmsVSfreq}(b)], MSD increases with the DCF only at lower frequencies; at higher frequencies, MSD actually falls with the DCF. The reason for this is that when the DCF is large compared to the hopping frequency, $\kappa/f_\text{D}\ll1$, on timescales larger than $1/\kappa$, the system behaves with the time-average of the disorder. In this regime the static components become increasingly significant with the DCF. For the defectless case this means the system becomes increasingly dispersive for larger DCF, as the static components are uniform. This was seen in \sref{sec:Frequency Controlled Diffusion}, where the system behaves increasingly dispersive with increasing DCF. In contrast, for the case with defects, the system increasingly becomes localised in the higher DCF regime, as the static components (specifically $\beta_r$) are highly disordered.

Also plotted in \fig{fig:rmsVSfreq}(a) (dotted line) is the MSD at time $T$, of the purely dispersive system with no defects (uniform $\beta$) and no dynamic disorder; plotted in \fig{fig:rmsVSfreq}(b) (dotted line) is the samples averaged MSD at time $T$ of the system with defects (stochastic $\beta_r$) and no dynamic disorder. \fig{fig:rmsVSfreq}(a) and (b) show that the MSD of systems with dynamic disorder do not approach that of the systems without dynamic disorder, even at high DCFs, which is qualitatively consistent with previous results for dynamic disorder with no temporal correlation ~\cite{madhukar77}.

\section{Conclusion}

We showed that stochastic time-dependent disorder can give rise to diffusive behaviour in JCH systems. We showed that the level of diffusivity can be controlled by the DCF. In the defectless case, MSD increases with DCF.  Interestingly however, when defects are present, MSD only increases with DCF at low frequencies;  at higher frequencies, MSD actually decreases with DCF, as  time-averaged behaviours set in. These behaviours arise from the time-dependent interplay between the intersite coupling and onsite repulsion. Therefore, although this work has concentrated on the JCH model as a specific case study, similar behaviour should also be present in other discrete model with similar properties, such as the Bose-Hubbard model. This work paves the way for diffusivity to be a controllable property in quantum metamaterials and quantum emulators.

\section{Acknowledgments}

The author would like to thank C.-H. Su for his feedback on the manuscript, A . L. C. Hayward and A. D. Greentree for fundamental conceptual discussions, and S. M. Quach for support and general discussions.

\bibliography{diffusion_pra}

\end{document}